# High-resolution neutron imaging: a new approach to characterize water in anodic aluminum oxides

Noémie Ott[a,*], Claudia Cancellieri[a], Pavel Trtik[b], Patrik Schmutz[a]

[a] Laboratory for Joining Technologies and Corrosion, Empa, CH-8600 Dübendorf, Switzerland
[b] Neutron Imaging and Activation Group, Laboratory for Neutron Scattering and Imaging, Paul Scherrer Institut, CH-5232 Villigen PSI, Switzerland

* Email: noemie.ott@empa.ch

**Highlights**

- High resolution neutron imaging was used to quantify water in anodic Al oxide
- Small changes in water spatial distribution can be detected by neutron imaging
- Water content is related to the anodizing parameters (electrolyte, anodizing time)
- Both structural (oxide) and morphological (pores) water contribution are identified
- Neutron imaging is sensitive to structural disorder and orientation of crystallites

**Abstract**

During the growth of anodic Al oxide layers water incorporates in the film and therefore influences the intrinsic properties of the oxide formed. In this study, we propose a new approach, based on the use of high-resolution neutron imaging, to visualize and quantify the water content in porous Al oxides as a function of anodizing conditions. Water in these porous films is either incorporated directly in the oxide structure (structural) and/or fills the pores (morphological). This preliminary study demonstrates that the differences in water content of porous anodic Al oxide layers are strongly related to the oxide growth parameters but interestingly cannot be directly correlated to a specific change in the amorphous oxide structure or in the pore morphology. Due to the high sensitivity of high-resolution neutron imaging to small changes in the water content, we furthermore show that the morphological water content in Al oxides formed in sulfuric acid as well as in phosphoric acid is partially reversible upon heat treatment and immersion. High-resolution neutron imaging is also found to be highly sensitive to structural disorder and crystallographic orientations, allowing to identify different crystalline Al oxide samples based on their structural and morphological defect content. This offers new perspectives to study the effect of the hydrogen and/or water incorporation as well as oxide-related structural modifications on Al oxihydroxides in relation to growth parameters, stability, functionalization as well as properties tuning, highly relevant to surface protection and their use as templates.

**Graphical Abstract**

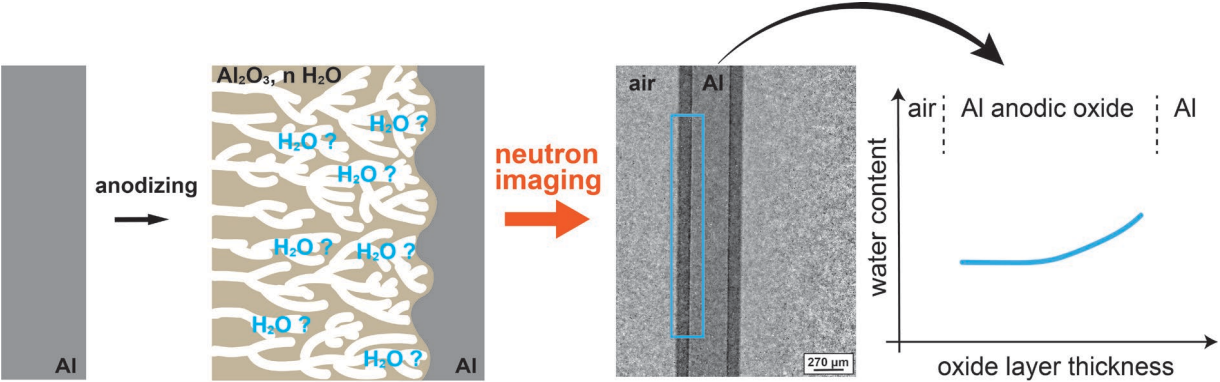

# 1. Introduction

Porous anodizing is a process widely used in industry as a surface finish for aluminum substrates to provide corrosion protection, wear resistance and decorative appearance [1-3]. It has also recently attracted great interest as a simple and inexpensive method to produce nanostructured materials [4-6]. By varying the anodizing conditions, the structure of the formed porous anodic Al oxide can be tuned and subsequently functionalized, making it highly attractive for various industrial applications from nanoscale electronics and optoelectronics to catalysts and sensors [4,5,7-14].

Porous anodic Al oxide layers consist of locally close-packed arrays of columnar hexagonal cells of anodic Al oxide arranged around a central pore and, separated by a barrier-type oxide layer from the metallic substrate [3,15,16]. The morphology of these layers (pore diameter, interpore distance and thickness of the barrier-type oxide layer) is determined by the anodizing parameters, including the electrolyte, temperature, anodizing regime (potentiostatic or galvanostatic) and duration [3,16-19]. Despite the apparent simplicity of the anodizing process, the formation mechanisms of porous anodic Al oxide layers are complex and still await full understanding. Most of the existing theories suggest that porous anodic Al oxide forms via field-assisted dissolution induced by the electrolyte aggressiveness [20-29]. Recently, an alternative model was proposed in which the pore formation during anodizing in phosphoric and sulfuric acids is associated to the field-assisted plasticity and species migration across the barrier-type oxide layer under growth stresses [30-37]. The electric field drives cooperative transport of cations ($Al^{3+}$, electrolyte species) and anions ($O^{2-}$, $OH^-$, electrolyte species), essential to film growth [16,20,22-24,38-41]. The porous anodic Al oxide thus formed mainly consists of amorphous, hydrated alumina with additional incorporation of electrolyte anions [31,35,38,41-46].

The presence of water related species, such as proton or hydroxyl groups integrated in the oxide structure or water of hydration, influences the growth and properties of porous anodic Al oxide layers [47-49]. It is therefore essential to include a proper quantification of these species when discussing the nature and structure of the formed anodic Al oxides. So far, the water content in anodic Al oxide layers has been estimated using AC bridge measurement [47,48,50], infrared spectroscopy (IR) [42], secondary ion mass spectroscopy (SIMS) [51-53], nuclear magnetic resonance (NMR) [54], $N^{15}$ nuclear reaction analysis (NRA) [55,56], elastic recoil detection (ERDA) [56], reflectivity techniques [57] and quartz crystal microbalance [58]. Although these techniques provide specific information about the hydrogen distribution as well as its chemical state, most of them do not allow the determination of water distribution profile in a micrometer range field of view and therefore are not suitable to investigate thick oxide layers.

The present study explores the applicability of high-resolution neutron imaging to characterize porous anodic Al oxides. Neutron imaging is a non-destructive technique, that is – due to the high sensitivity of neutrons to hydrogen – very appropriate for assessing the distribution of hydrogen-containing compounds, such as water, in structural materials that are almost transparent to neutrons (e.g. those based on aluminum). Thus, neutron imaging has been successfully used in a variety of research areas to quantify the water content in natural and engineered porous materials, such as rocks, soils, concrete and fuel cells [59-66]. To our best knowledge, it has not yet been used to study anodic oxides due to its limited spatial resolution of several tenths of micrometers. To overcome this limitation, a high-resolution imaging setup, the neutron microscope detector, has been developed at Paul Scherrer Institute (PSI) in Switzerland [67]. This state-of-the-art setup provides an unprecedented spatial resolution below 5 μm at an effective pixel size of 1.3 μm [67-70].

The purpose of this study is therefore to address the prospects of high-resolution neutron imaging to characterize porous anodic Al oxides. With this in mind, we have investigated the effect of substrate purity, anodizing electrolyte and anodizing time on the hydrogen integration and water content of the formed porous anodic Al oxides. The findings provided by neutron imaging were correlated to the structural and morphological evolution of the oxide layers. As few fundamental neutron studies exist, the results about anodic oxides were further compared to the one obtained for other typical Al oxide samples, such as crystalline C-sapphire, sintered Al oxide, plasma sprayed Al oxide and an Al hydroxide, Al(OH)$_3$, representing various other structures and defect/water distribution.

## 2. Materials and methods

### 2.1. Samples

This study focuses on porous Al oxide layers obtained by galvanostatic anodizing of pure Al substrates. Two Al purity grades, pure Al (Al 99.9%, 0.5 mm thick, Novelis, Switzerland) and AW1050 (Al 99.5%, 1.1 mm thick, Novelis, Switzerland), were investigated. To meet the requirements of the neutron imaging setup, the Al substrates were cut as following: about 2.00 mm in depth (plate width placed parallel to the beam) – an actual depth of 2.05 ± 0.03 mm was measured – and 25 mm in height. The substrates were used in the as-received condition. Prior to anodizing, the samples were first ultrasonically cleaned for 5 min in acetone, then for 5 min in ethanol and finally dried with Ar.

Double-sided galvanostatic anodizing was carried out at a current density of 35 mA cm$^{-2}$ and at room temperature in either 0.1 M phosphoric acid or 0.5 M sulfuric acid, both unstirred. A two electrode electrochemical cell was used, the sample acting as the working electrode and a Pt ring electrode as counter electrode. A Keithley 2400 SourceMeter SMU instrument (Tetronix) was used to perform the anodizing process. The anodizing times for pure Al were, respectively, 2320 s, 6900 s and 9300 s in 0.1 M H$_3$PO$_4$ and 3000 s in 0.5 M H$_2$SO$_4$. For comparison purposes, AW1050 was anodized for 6700 s in 0.1 M H$_3$PO$_4$. After anodizing, the samples were rinsed with MilliQ water (Merck, 18.2 MΩ cm) and dried with Ar. The anodizing times were calculated, based on a preliminary study, to form oxide layers with respective thicknesses of 50 µm, 150 µm (both on pure Al and AW1050) and 250 µm in 0.1 M H$_3$PO$_4$ and of 50 µm in 0.5 M H$_2$SO$_4$. Table 1 summarizes the anodizing conditions for the different samples and the effective thicknesses obtained.

As a comparison, the following Al oxides were also investigated: C-sapphire ([0001] oriented, Crystec), sintered Al oxide (99.6% purity, Kyocera), plasma sprayed Al oxide (Nova Swiss) and a commercial Al oxide membrane (Anodisc 13, Whatman®). Additionally, a 50 µm thick Al hydroxide layer was prepared by evaporating a saturated Al(OH)$_3$ (gibbsite) solution on a pure Al substrate. The samples were cut to present a sample depth of 2.00 ± 0.05 mm in the beam direction.

### 2.2. Microstructure and structure characterization

The thickness and morphology of the different oxide layers were studied using a FEI Nova NanoSEM field emission gun scanning electron microscope (FEG-SEM). Cross-sections were prepared using a Hitachi IM4000 Ar ion milling system. All samples were coated with a 5 nm conductive gold layer prior to SEM investigation.

Structural characterization of the anodic Al oxides and of the reference samples was performed via X-ray diffraction (XRD) in a Bruker D8 diffractometer in Bragg Brentano geometry using Cu Kα radiation and a Ni filter.

### 2.3. High-resolution neutron imaging
#### 2.3.1. Neutron microscope setup and operation conditions

High-resolution thermal neutron imaging was performed using the neutron microscope at the Pulse OverLap Diffractometer (POLDI) beamline at the Swiss spallation neutron source (SINQ), PSI. Spectral information are reported in [71]. The samples were measured during two different neutron beamtimes. Anodized Al layers, C-sapphire and Al hydroxide samples were measured using the detector being equipped with a sCMOS camera with the resulting image pixel size of 2.6 µm. The sintered Al oxide, plasma sprayed Al oxide and commercial Al oxide membrane were measured using the detector being equipped with a CCD camera with the resulting image pixel size of 2.7 µm (for details, see supplementary information). C-sapphire has then been measured in both detector configurations for comparison purposes. The state-of-the-art neutron microscope setup was equipped with a 3.5 µm thick 157-gadolinium oxysulfide scintillator and in both cases, the samples were placed at about 2.5 mm from the scintillator. The setup therefore provides an achievable spatial resolution of at least 9.6 µm [67,70,72,73]. Figure S1 shows the setup used for these experiments.

For each sample, a stack of 100 projection images was acquired with an exposure time of 30 s per projection. The black body correction procedure developed at PSI [74,75] was applied to allow for quantitative analyses. Therefore, stacks of at least 25 open beam (OB) – images and open beam images with an interposed frame containing neutron absorbers, black bodies (BB), as well as a stack of 100 images of sample with the BBs were additionally taken. The exposure time was 30 s per radiograph. The mentioned stacks of images were averaged over the entire stack depth to form a single image. After having been corrected for scattering and bias, the obtained image $T$ was used for quantitative analyses (Figure S2). The detailed data processing is described in the supplementary information.

### 2.3.2. Quantifying the water content

Since the attenuation of radiation passing through matter follows the Beer-Lambert law, the transmission contrast $T$ can be expressed as:

$$T = e^{-\Sigma_{tot} d} \quad (1)$$

where $\Sigma_{tot}$ is the linear attenuation coefficient of the material [cm$^{-1}$] through $d$, the sample depth in the beam direction equal to 2.05 ± 0.03 mm for all studied samples.

Regardless of its structural composition, it is generally assumed that porous anodic Al oxide mainly consists of hydrated Al oxide, referred as $Al_2O_3 \cdot n\ H_2O$. This water comprises the contribution of both the water/hydrogen incorporated in the oxide structure and the hydration water that remains in the pores. The contribution of the electrolyte anions adsorbed at the anodic Al oxide/electrolyte interface to the transmission contrast is neglected because both sulfur and phosphorus possess rather low neutron cross sections. The transmission contrast $T$ can consequently be expressed as follow:

$$T = e^{-(\Sigma_{porous\ Al_2O_3,dry}\ d\ +\ \Sigma_{H_2O}\ d_{H_2O})} \quad (2)$$

where $\Sigma_{porous\ Al_2O_3,dry}$ is the linear attenuation coefficient of water-free porous anodic Al oxide [cm$^{-1}$], $\Sigma_{H_2O}$ is the free water linear attenuation coefficient [cm$^{-1}$], $d$ is the sample depth in the beam direction equal to 2.05 ± 0.03 mm for all the studied samples and $d_{H_2O}$ is the equivalent water thickness [cm], which includes the contribution of the water/hydrogen incorporated in the oxide structure (structural water) and/or the hydration water that remains in the pores (morphological water). Note that Eq. 2 is an approximation of the real system, since the contribution of structural and morphological water cannot yet be directly distinguished with neutron imaging.

The overall water content of the anodic Al oxide layers, described by the equivalent water thickness $d_{H_2O}$, can be deduced from Eq. 2:

$$d_{H_2O} = -\frac{\ln(T) + \Sigma_{porous\ Al_2O_3,dry}\ d}{\Sigma_{H_2O}} \quad (3)$$

provided that the neutron cross-sections are known. Those can be calculated from the atomic ones but, in the case of a multi-element compound, the empirical and calculated values differ due to non-negligible additional scattering effects (Table S1). We therefore need to rely on experimental referencing to determine the linear attenuation coefficient.

The free water linear attenuation coefficient at the POLDI beamline, $\Sigma_{H_2O}$, has been previously measured and is equal to 3.78 cm$^{-1}$. Since finding a suitable reference sample for the porous anodic Al oxides proved to be challenging, the following approaches were taken to determine the cross-section, i.e. linear attenuation coefficient of "dry" porous anodic $Al_2O_3$. Anodized samples were placed in ultra-high vacuum (UHV – ~ 10$^{-8}$ mbar) for 12 h in an attempt to desorb water from the oxide layers. On the other hand, selected anodized samples were heated for 1 h (H1), 3 h (H3) and 5 h (H5) at 150 °C on a hot plate. The temperature was chosen to induce as little change to the oxide as possible. The "dried" porous oxide was then immersed for 30 min in water to determine if the pores could be refilled with water. After each step, neutron imaging was performed in order to monitor the water content evolution in the porous anodic films. The highest contrast transmission value obtained, corresponding to "dry" $Al_2O_3$, was used as reference value to derive the linear attenuation coefficient of water free porous anodic $Al_2O_3$. The water content determined is therefore relative to this "dry" $Al_2O_3$ value.

## 3. Results
### 3.1. Microstructure and structure of porous anodic Al oxide layers

In this study, one-step galvanostatic anodizing at a current density of 35 mA $cm^{-2}$ was performed. Anodizing pure Al in 0.1 M $H_3PO_4$ for 2320 s, 6900 s and 9300 s, results in the formation of porous anodic Al oxide layers with respective thicknesses of 50 μm (pure Al P50), 130 μm (pure Al P130), and 230 μm (pure Al P230) whereas only a 85 μm thick anodic Al oxide layer (1050 P85) was formed on AW1050 substrate anodized for 6700 s in 0.1 M $H_3PO_4$, as determined by SEM. Anodizing pure Al in 0.5 M $H_2SO_4$ for 3000 s at 35 mA $cm^{-2}$, on the other hand, leads to anodic Al oxide films with a thickness of around 55 μm (pure Al S60), see Table 1.

Table 1: thickness and pore size of the different porous anodic Al oxides investigated.

|  | electrolyte | anodizing time [s] | thickness – SEM [μm] | thickness – neutron images [μm] | pore size [nm] |
|---|---|---|---|---|---|
| pure Al S60 | 0.5 M $H_2SO_4$ | 3000 | 62 ± 6 | 54 ± 3 | 20-25 |
| pure Al P50 | 0.1 M $H_3PO_4$ | 2320 | 50 ± 1 | 45 ± 3 | 220 ± 40 |
| pure Al P130 |  | 6900 | 130 ± 1 | 122 ± 6 | 220 ± 40 |
| pure Al P230 |  | 9300 | 217 ± 5 | 238 ± 10 | 220 ± 40 |
| 1050 P80 |  | 6700 | 80 ± 3 | 83 ± 3 | 230 ± 30 |

Figure 1 reports typical voltage evolution related to growth mechanisms as a function of anodizing time for the different samples. During the initial stage of anodizing, regardless of the substrate or electrolyte, a steep increase of voltage, associated to the growth of a barrier-type oxide layer, is observed for all samples. The voltage then reaches a maximum when field-assisted chemical dissolution processes take place and pores start to form. It is followed by a voltage decay to a plateau, which corresponds to the steady-state growth of the porous oxide layer. After long anodizing times in 0.1 M $H_3PO_4$ (Figure 1b and c), the voltage starts increasing again. This increase was reproducible for a given anodizing condition, as shown by the curves of the two replicates of pure Al P130 R1 and R2 (R indicating the replicate).

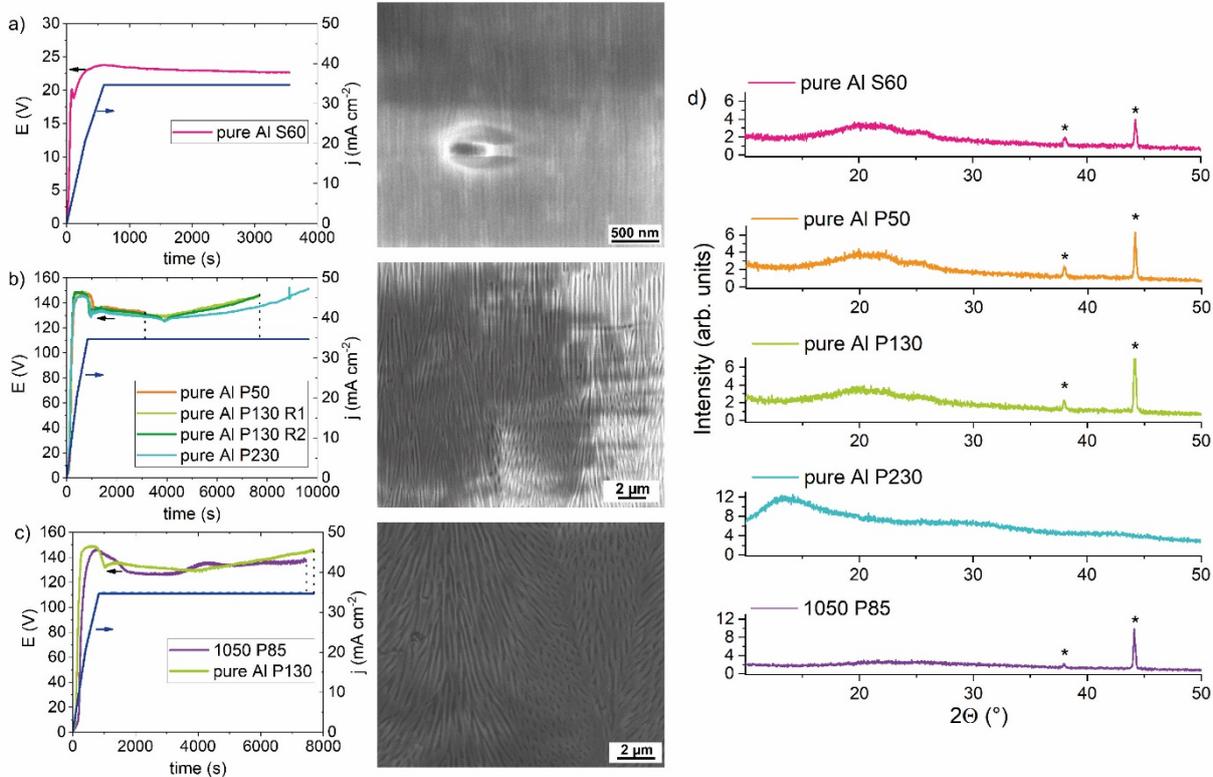

Figure 1: Typical anodizing curves and SEM cross section images of the resulting porous anodic oxide layers obtained for a) pure Al anodized in 0.5 M $H_2SO_4$; b) pure Al anodized in 0.1 M $H_3PO_4$ and c) AW1050 anodized in 0.1 M $H_3PO_4$ with pure Al P130 as comparison. Pure Al P130 R1 and R2 are two replicates of the same anodizing conditions. d) Corresponding XRD scans. The peaks marked with * are coming from the Al substrate.

There are notable differences in the voltage values measured depending on the anodizing electrolyte used, as would be expected, which induces significant differences in the structure of the porous anodic Al oxide layer. Under the anodizing conditions selected in this study, anodizing pure Al in 0.5 M $H_2SO_4$ leads to the formation of a very defective barrier-type oxide, reflected by the low voltage values (around 20 V) reached during growth (see Figure 1a). SEM observations of the porous Al anodic oxide layer in cross sections show long-range ordered pores, perpendicular to the growth direction through the whole oxide thickness (see Figures 1a and S3). Some imperfections in the pore arrangement can nevertheless be observed in Figure 1a. The pore diameter was estimated to be around 20-25 nm (see Table S2).

During anodizing in 0.1 M $H_3PO_4$, higher voltage values in the range of 130-140 V are reached, as shown in Figure 1b and c, indicating that the barrier-type oxide formed is less defective than in sulfuric acid. However, compared to the films grown in sulfuric acid, the porous anodic Al oxide layers formed in phosphoric acid present a lower degree of pore ordering. Considering the anodic oxide layers formed on pure Al, the top part of the layer exhibits self-ordering perpendicular to the growth direction (see Figures 1b and S5). The thickness of this ordered layer is 20-25 µm, regardless of the anodizing time, which suggests that the pore order could only be maintained during the initial stages of anodizing. Prolonged anodizing leads to random pore arrangement and structural disorder associated with the formation of interconnected, branched and terminated pores. Despite the loss of long-range pore order, short-range domains of ordered pores develops, as evidenced in Figure S5. This is also observed for the porous anodic Al oxide layers grown on AW1050 but interestingly, no self-ordering occurs for this substrate (see Figure 1c and S6). Different pore ordering does not seem to influence the pore size, which is found to be around 220-230 nm on both substrates (see Table S2). It is however worth mentioning that identical current density and similar potential evolution lead to the formation of an 85 µm-thick oxide layer on AW1050 compared to 130 µm on pure Al, which suggests a lower efficiency of the oxide layer growth on AW1050 substrates.

From the XRD analysis presented in Figure 1d, it can be deduced that all the porous Al anodic oxides formed are in an amorphous state. This statement is supported by the absence of sharp diffraction peaks and by the presence of broad reflections between 10 ° and 30 °. However, small crystallites (nanocrystals) inside the oxide could also result in such diffractograms, thus their presence cannot be completely excluded. Interestingly, as shown in Figure 1d, the position and width of these broad XRD reflections (related to the presence of local structured areas) are similar for pure Al S60, pure Al P50 and pure Al P130. Their "amorphous" XRD fingerprint shows three broad intensity reflections around the 2θ values of 20 °, 22 ° and 25 °, which can be attributed to $Al(OH)_3$ (gibbsite)-, $AlO(OH)$ (diaspore)- and $\alpha$-$Al_2O_3$ (corundum)-like structures, respectively (see Figure S8). This is a clear indication that Al anodized layers cannot be solely described as Al oxides but consist of a complex arrangement of Al oxide, oxihydroxides and hydroxides. For pure Al P230, on the other hand, a shift of the amorphous bump towards lower 2θ values is observed, which indicates a structural change in the oxide associated to the presence of $AlO(OH)$ (boehmite)-like structures, instead of $Al(OH)_3$. This structural change for thick porous anodic Al oxides can be related to an increase of the temperature of the anodizing bath over time due to resistive heating associated to the high voltage values reached during anodizing in phosphoric acid, since the temperature was not controlled. Although hardly distinguishable, the XRD pattern of 1050 P85 is similar to pure Al P130, with the exception of the amorphous bump around 20 °, suggesting that gibbsite is not formed (see Figure S8). The lower intensities and broader peaks observed in the diffractogram of 1050 P85 are furthermore an indication of a higher structural disorder of the Al oxide, compared to the anodic films grown on pure Al. Consequently, the XRD investigation shows that the porous anodic Al oxide layers formed exhibit morphological and/or intrinsic structural differences depending on the anodizing conditions.

## 3.2. Neutron imaging – transmission contrast profiles

High-resolution neutron imaging was performed on the different samples to first evaluate their water content (based on hydrogen content detection) as a function of anodizing conditions. Using the contrast differences observed (see Figure 2), the thickness of the porous anodic Al oxide layers can be directly derived from the obtained transmission contrast images. The values found are in good agreement with the values retrieved from SEM cross section images, as reported in Table 1. The remaining thickness of the substrate is also easily accessible.

To allow a more detailed comparison between the different anodized samples, transmission profiles are extracted from the neutron radiographs, as indicated in Figure 2. Within the spatial resolution of neutron imaging, the samples can be considered homogeneous in the z direction (see Figure 2a). The extracted profiles therefore display the transmission values averaged in the z direction and plotted over the sample width within the selected area. For comparison and referencing purposes, the profiles were always plotted with the air on the left and the metallic substrate on the right. The zero of the x axis was arbitrary set at the half maximum of the air/sample interface. Figure 2 provides typical transmission images for pure Al, anodized pure Al (pure Al P130 H5), C-sapphire and Al(OH)$_3$ and their corresponding extracted contrast transmission profiles. Despite being visually different on the images of Figure 2 due to different contrast scales, the transmission value $T$ for air is 1.00 for all samples.

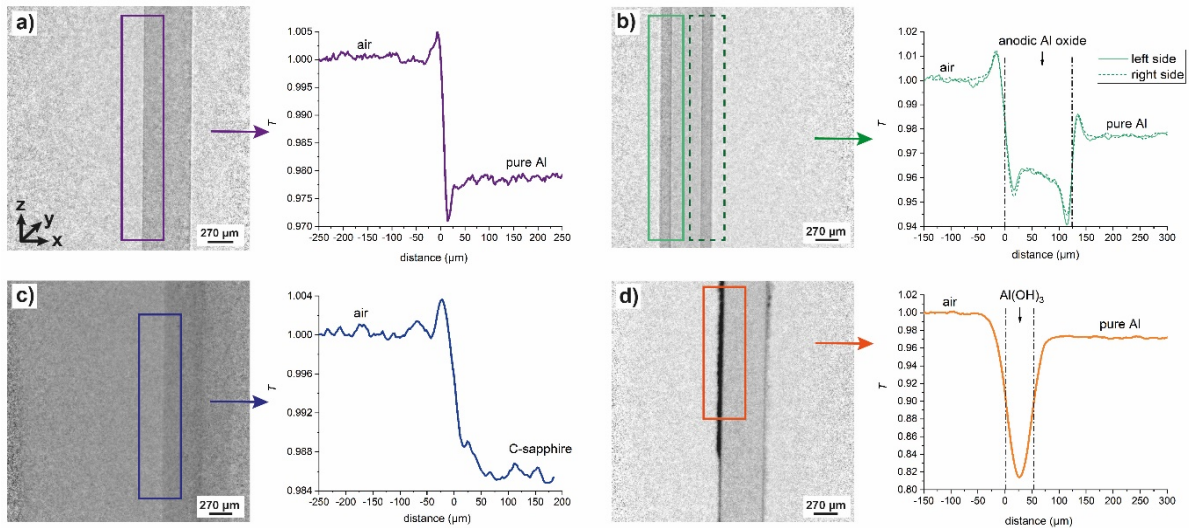

Figure 2: Transmission contrast images $T$ and their corresponding extracted profiles for a) pure Al, b) porous anodic Al oxide (pure Al P130 H5), c) C-sapphire and d) Al hydroxide, Al(OH)$_3$. In those profiles, the contrast transmission values are averaged in the z direction and plotted over the sample width (through the selected area). The zero of the x axis is arbitrary set at the half maximum of the air/sample interface. The neutron beam is in the y direction.

An important point that needs to be mentioned concerning neutron imaging is the presence of edge enhancement effects at the interfaces. The transmission profiles of Figure 2 clearly show that none of the interfaces is sharp and that significant deviations in the neutron transmission values, either enhancement or drop, are observed at the different interfaces (air/oxide and oxide/Al metallic substrate). This effect can partly be attributed to the sample roughness and to an imperfect parallel positioning of the sample compared to the beam. However, it is mostly intrinsic to neutron imaging and is related to additional refraction and reflection of the neutrons at the materials interfaces [76-78]. These edge enhancement effects currently limit data quantification in the affected interface regions – about 25 µm from both sides of each interface. Consequently, the neutron data discussed in this paper exclude those regions. Moreover, due to the extent of the edge enhancement effects, the minimum thickness (oxide and substrate) that can be investigated in this study is therefore about 50 µm, as shown in Figure 2d. The Al hydroxide sample has been prepared by evaporating a saturated Al(OH)$_3$ solution on a pure Al substrate, which allows to obtain such a thick hydroxide layer (left side in Figure 2d) but the whole sample is covered by a thin inhomogeneous hydroxide layer, which accounts for the dark contrast at the air/sample interfaces. This particular sample preparation results in rather rough interfaces, which contribute to broaden the interface region.

Figure 3 presents the transmission profiles obtained for the different porous anodic Al oxide layers investigated. The full profiles are reported in Figure S9. The pure Al substrate, bare or anodized, has a transmission value of 0.978 (see Figures 2 and S9), except for pure Al P230, for which the transmission value lessens to 0.957. For this sample, the remaining substrate is very thin, due to the double-sided anodic growth of around 230 µm thick oxide layers. It is hypothesized that during anodizing, due to the metal thinning, crystalline defects are introduced in the substrate, resulting in this lower transmission value. As for AW1050 substrate, its transmission value of 0.978 is comparable to pure Al, which indicates that the small additional amount of alloying elements does not influence the neutron transmission value. The transmission values of 1.00 for air and 0.978 for Al substrates therefore act as reference values when discussing the changes in contrast transmission values for the porous anodic Al oxide layers. The two sides of an anodized sample are also expected to present identical profiles, as illustrated in Figure 2b.

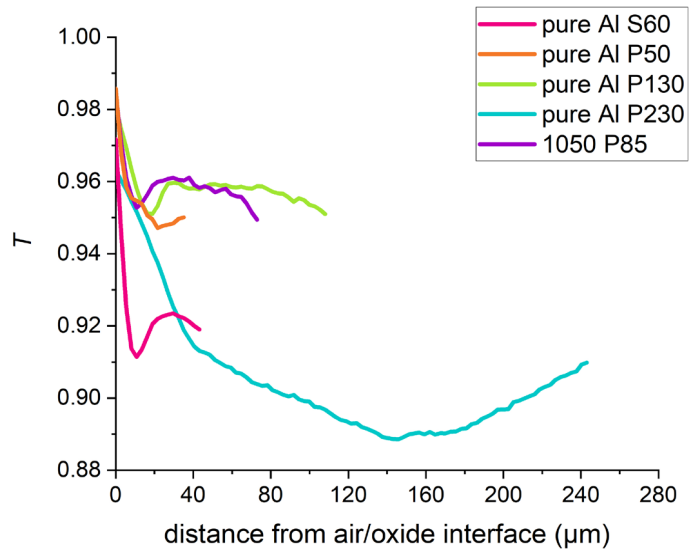

Figure 3: Typical contrast transmission profiles for pure Al anodized in 0.5 M $H_2SO_4$ for 3000 s (pure Al S60), in 0.1 M $H_3PO_4$ for 2320 s (pure Al P50), 6900 s (pure Al P130) and 9300 s (pure Al P230) and AW1050 anodized in 0.1 M $H_3PO_4$ for 6700 s (1050 P85) at 35 mA cm$^{-2}$. The oxide/metallic substrate interface has been removed to ease the comparison.

As reported in Figure 2, the contrast transmission value of single crystal C-sapphire, dense oriented crystalline $Al_2O_3$, is 0.988. As would be expected, the transmission value for all porous anodic Al oxides is lower than the transmission value of single crystal corundum. Since the electrolyte species are neutron transparent, the lower values and changes in the transmission value obtained for the anodized layers can be solely attributed to the difference in water content. Due to the dominating influence of the hydrogen neutron cross section, it can be postulated that the lower the transmission value, the more hydrogenous species are contained in the porous films, either as structural (incorporated in the oxide) and/or morphological (in the pores) water. Figure 3 shows moreover that the transmission value seems to strongly vary depending on the anodizing conditions. Variation of the transmission value within each individual sample is observed, suggesting changes in the water content across the oxide thickness. Pure Al P130 has a contrast transmission value of 0.959 near the air/oxide interface but the value lowers towards the oxide/Al substrate interface, which indicates a water enrichment deeper in the oxide near the interface with the metal. Pure Al P50 and P230 exhibit a different behavior compared to pure Al P130. Their transmission value is higher at both interfaces and presents a minimum in the middle of the oxide layer. Pure Al P50 has a transmission value around 0.95, which means that it is slightly more hydrated than pure Al P130, although both present the same pore morphology and amorphous oxide structure (Figure 1). On the other hand, the porous Al anodic oxide layer (pure Al P230) formed on pure Al after 9300 s anodizing in 0.1 M $H_3PO_4$ shows the lowest transmission values ranging from 0.92 near the interfaces to 0.89 in the middle of the layer. Since all the anodized layers in phosphoric acid have similar pore dimensions and, we assume similar porosity, a sole increase of water intake from the pores cannot explain the difference in the transmission value between pure Al P230 and pure Al P130 while it can be attributed to the oxide structural changes observed in the XRD diffraction pattern, associated to the presence of boehmite-like structure for Al P230 instead of gibbsite-like for Al P130 (Figure S8).

Interestingly, at equivalent thickness, the porous anodic Al oxide formed in sulfuric acid is more hydrated than the one formed in phosphoric acid, as shown by the transmission value difference in Figure 3. Although both types of anodic oxides present a similar amorphous oxide structure, they are morphologically different, i.e. they have a different pore structure and are expected to present a different composite (oxide and pore) density. The composite density for porous anodic Al oxides formed in sulfuric acid is expected to be slightly lower than the ones formed in phosphoric acid, which would lead to a higher transmission value if both oxides contained the same water content. The lower transmission value measured for pure Al S60 shows that the long-range arrangement of 20 nm pores that constitutes pure Al S60 seems to favor water entrapment in the pores compared to pure Al P50, especially towards the oxide/Al substrate interface.

Considering the porous anodic Al oxide formed on 1050, its transmission values are similar to pure Al P130, except for a higher contrast near the oxide/Al substrate interface. This decrease in the transmission value towards the substrate can be related to more pore disorder and possibly more pore blockage near the oxide/AW 1050 substrate interface, favoring water entrapment compared to pure Al anodized layers.

These results evidence that the water content in the porous anodic Al oxides is related to structural water, i.e. water/hydrogen incorporated in the oxide structure, and to morphological water, i.e. water filling the pores. The amount of morphological and structural water strongly depends on the anodizing parameters and subsequently, on the oxide grown. Nevertheless, no straightforward correlation could be established between the pore morphology and oxide structure and the water content of the porous anodic Al oxide layers. Pure Al P50 and pure Al P130 present the same pore and oxide structures and still, exhibit a different contrast transmission profile. Including the water content in the discussion about the nature and structure of anodic oxides is therefore essential.

### 3.3. Quantification of water in porous Al oxide layers

The contrast transmission profiles provide valuable qualitative information about the morphological and structural water content and distribution in porous anodic Al oxide layers. In a next step, Eq. 3 was used to quantify this water content. Since producing a water-free anodic Al oxide that can act as reference to determine the linear attenuation coefficient of "dry" porous anodic $Al_2O_3$ proved to be challenging, one pure Al P130 sample was stored for 12 h in ultra-high vacuum (pure Al P130 UHV) while another was heat-treated, in an attempt to desorb the water initially present in the pores and subsequently, obtain a water-free porous $Al_2O_3$. In this preliminary study, pure Al P130 was selected as a quasi-reference sample, due to its initial low morphological water content and similar oxide structure compared to pure Al P50 and 1050 P85 (Figure 3). We are nevertheless aware that it remains a quasi-reference sample, since it presents structural differences with pure Al P230 and morphological differences with pure Al

S60, which can lead to over- and underestimation of the water content in these two samples, respectively.

The heat treatment was performed at 150 °C on a hot plate for 1 h (pure Al P130 H1), 3 h (pure Al P130 H3) and 5 h (pure Al P130 H5). It was then immersed in water at room temperature for 30 min (pure Al P130 W) to determine if the pores could be refilled with water. Neutron imaging was performed after each step to monitor the evolution of the morphological water content in the porous anodic films. The contrast transmission profiles of these samples are displayed in Figure 4a.

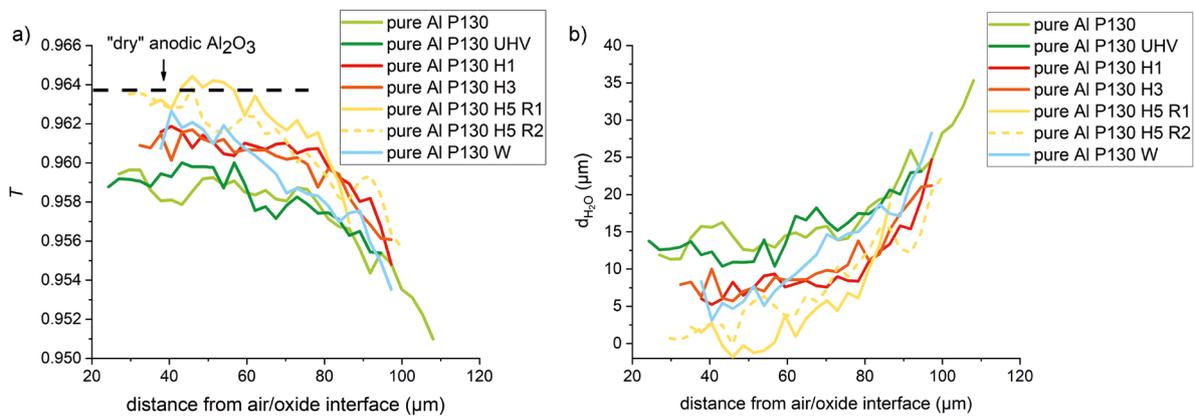

Figure 4: a) Typical contrast transmission profiles and b) equivalent water thickness, $d_{H_2O}$, for pure Al P130 heated for 1 h (H1), 3 h (H3) and 5 h (H5) at 150 °C and followed by immersion in water at room temperature for 30 min as well as pure Al P130 UHV, let for 12 h in ultra-high vacuum. The profile of pure Al P130 is given as an initial reference. Pure Al P130 H5 R1 and R2 are two replicates of the same sample conditions. The equivalent water thickness is expressed for the total sample depth of 2.05 mm.

Pure Al P130 UHV presents a similar profile to pure Al P130, with a maximum $T$ value of 0.959 near the air/oxide interface, indicating that ultra-high vacuum is not efficient to desorb the water from the porous films. On the other hand, heating the sample leads to higher contrast transmission values in the top part of the oxide layer (Figure 4a).

Since no changes were observed in the oxide structure (Figure S8) or in pore morphology after heating, this increase can be attributed to morphological water desorption occurring in the pores during the thermal treatment. The highest contrast transmission value was obtained after 5 h of heating, with the transmission value reaching 0.964 compared to 0.959 for pure Al P130 near the air/oxide interface, as indicated by the dashed line in Figure 4a. This evolution is observed on both replicates, pure Al P130 R1 and R2. This value of 0.964 was assimilated to the transmission value of "dry" porous anodic and used to derive the linear attenuation coefficient, $\Sigma_{porous\ Al_2O_3,dry}$. It was found equal to 0.176 cm$^{-1}$, value that was used in Eq. 3 to quantify the water content of the porous anodic Al oxide layers. The deduced water content, described as the equivalent water thickness, $d_{H_2O}$, is reported in Figure 4b. It is necessary to mention that the quantified water content is relative to this specific "dry" $Al_2O_3$ value. It would be interesting to determine if all the morphological water can be removed by heating the sample without changing the oxide structure or pore morphology of the porous anodic Al oxide but this was not possible during the neutron beamtime allocated and therefore remains outside the scope of this paper.

Heating pure Al P130 for 1 h or 3 h leads to an increase of the transmission value from 0.959 to 0.961 that is equal to the desorption of 6 µm of water at top part of the oxide layer whereas after 5 h of heating, a water thickness of about 13 µm can be removed, which corresponds to about 0.73 % of the total oxide sample depth of 2.05 ± 0.03 mm. Interestingly, water desorption only occurs in the top first 70 µm of the oxide layer. No changes is detected in the profile near the oxide/Al substrate interface, suggesting water entrapment in the pores at the bottom part of the oxide, probably due to pore branching and subsequent blockage. This trend is observed for all heated samples, as shown in Figure 4. When pure Al P130 H5 is then immersed in water at room temperature for 30 min, the equivalent of 5.5 µm of water is adsorbed in the first 70 µm of the oxide layer, which shows that the pores can reversibly be filled with water. As before, the profile at the oxide/Al substrate interface remains unaffected, indicating that water cannot reach the pores at the bottom part of the oxide.

Figure 4 clearly demonstrates the potential of high-resolution neutron imaging to study the evolution of the water content and its distribution in porous anodic Al oxides. Small changes in the equivalent water thickness, up to a couple of µm, which can drastically influence the oxide properties and stability, can be easily detected with this technique while being highly challenging to observe using techniques involving photon or electron beams.

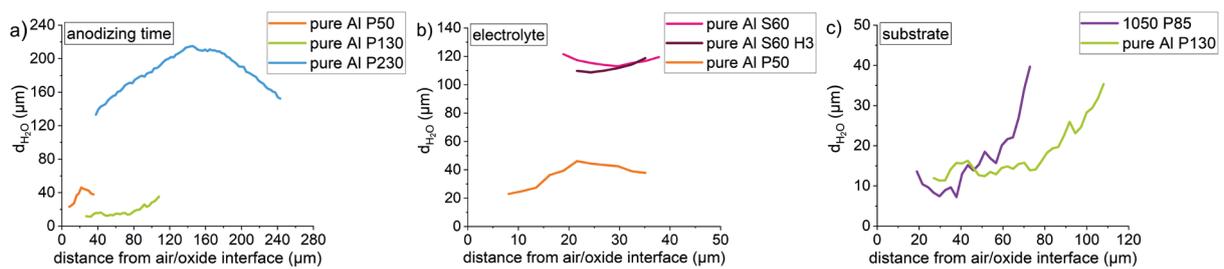

Figure 5: Equivalent water thickness, $d_{H_2O}$, for the different porous anodic Al oxide layers investigated a) effect of the anodizing time – pure Al anodized in 0.1 M $H_3PO_4$; b) effect of the anodizing electrolyte – pure Al anodized in 0.5 M $H_2SO_4$ compared to pure Al P50; and c) influence of the Al substrate –1050 P85 compared to pure Al P130. The equivalent water thickness is expressed for the total sample thickness of 2.05 mm. Note that the water content is relative to the "dry" porous anodic $Al_2O_3$ value.

Figure 5 displays the overall water content of the porous anodic Al oxide layers as a function of anodizing conditions. Note that the equivalent water thickness determined is relative to the "dry" porous anodic $Al_2O_3$ value. At the air/oxide interface, pure Al P50 contains twice as much water as pure Al P130, as reported in Figure 5a. Deeper in the oxide, the equivalent water thickness shows an increase of 25 µm followed by a loss of 10 µm towards the oxide/Al substrate interface. This increase occurs in the first 25 µm of the oxide layer, which corresponds to the thickness of the top ordered layer (Figure S5). This observation suggests a higher water intake in ordered pores compared to disordered, branched pores. This surface component is hidden in pure Al P130 due to edge enhancement effects while not distinguishable in pure Al P230 profile probably due to imperfect parallel positioning of the sample compared to the beam. Regarding pure Al P230, the structural change in the oxide associated to the presence of boehmite-like structure instead of gibbsite-like structure results in a water content ten times higher than pure Al P130. The corresponding water amount represents 7 to 11 % of the total sample thickness. Due to the structural differences between pure Al P130, used as quasi-reference sample, and pure Al P230, the water content of this sample may be overestimated but remains anyway higher than in pure Al P130. Similarly, the water content of porous anodic Al oxide formed in sulfuric acid may be underestimated because of their morphological differences. Therefore, at equivalent oxide thickness, Figure 5b indicates that porous anodic Al oxides formed in sulfuric acid contain at least two and half times more water than the ones formed in phosphoric acid. Pure Al S60 was then heated for 3 h at 150 °C, resulting in desorption of morphological water equivalent to 8 µm within the top first 30 µm of the oxide layer, which is comparable to the amount removed from pure Al P130 H3. Moreover, once again, no changes is observed in the profile near the oxide/Al substrate interface, implying that the water remains trapped in the pores at the bottom part of the oxide. This suggests that despite the long-range pore order observed in porous anodic Al oxides formed in sulfuric acid, water desorption from the pores is also only possible in the top part of the oxide layer as will probably be reversible filling of the pores. When using AW 1050 as a substrate instead of pure Al, the water content of the respective porous anodic Al oxides is comparable near the air/oxide interface, as shown in Figure 5c, whereas close to the oxide/Al substrate interface, the oxide formed on AW1050 contains two times more water than pure Al P130.

### 3.4. Neutron imaging of Al oxides and hydroxide reference samples

Additional high-resolution neutron imaging was performed on reference Al oxides – C-sapphire, sintered Al oxide, plasma sprayed Al oxide, a commercial Al oxide membrane – and an Al hydroxide, Al(OH)$_3$ to further explore the applicability of the technique for characterizing Al oxides. Although it would have been interesting to compare the previous findings with sub-micrometer thin, dense and compact Al oxides, such as native or barrier anodic Al oxides, they could not be included in the current study due to the current resolution of neutron imaging requiring films thicker than 50 µm.

Compared to the porous Al anodic oxide layers (Figure 2b), these samples are expected to be homogeneous within the resolution of the utilized neutron imaging detector. This is reflected in their transmission values, which remain constant over the image width in the x direction as shown in Figure 2a, c and d. The transmission value for each sample is therefore averaged over the sample width and is reported in Figure 6. The value of the C-sapphire is furthermore the average of the transmission values measured at the two beamtimes.

Figure 6 clearly evidences the differences in neutron transmission values between the Al oxides, depending on their production processes. As previously mentioned, the single crystal C-sapphire ([001] oriented perpendicular to the neutron beam) has a transmission value of 0.988, which is close to air. This suggests that dense, oriented single-crystalline alumina is almost neutron transparent. Like C-sapphire, the sintered aluminum oxide consists of corundum ($\alpha$-$Al_2O_3$) but is isotropic and polycrystalline. The resulting transmission value is around 0.955, much lower than for C-sapphire. This can be explained by diffraction contrast in transmission neutron imaging [79], whose transmission value depends on the structural crystalline features (planes, grains, orientation, number of crystallites), due to their effect on the material coherent elastic scattering cross-section. Thus, the sintered Al oxide presents a higher density of randomly oriented crystallites in the neutron beam compared to the single crystal sapphire, which induce stronger coherent elastic scattering and subsequently, result in a lower transmission value. The value measured for polycrystalline $\alpha$-$Al_2O_3$ (sintered Al oxide) is in good correlation with the value calculated using NEA Java-based Nuclear Data Information System (JAVIS)[80], see SI section 3.

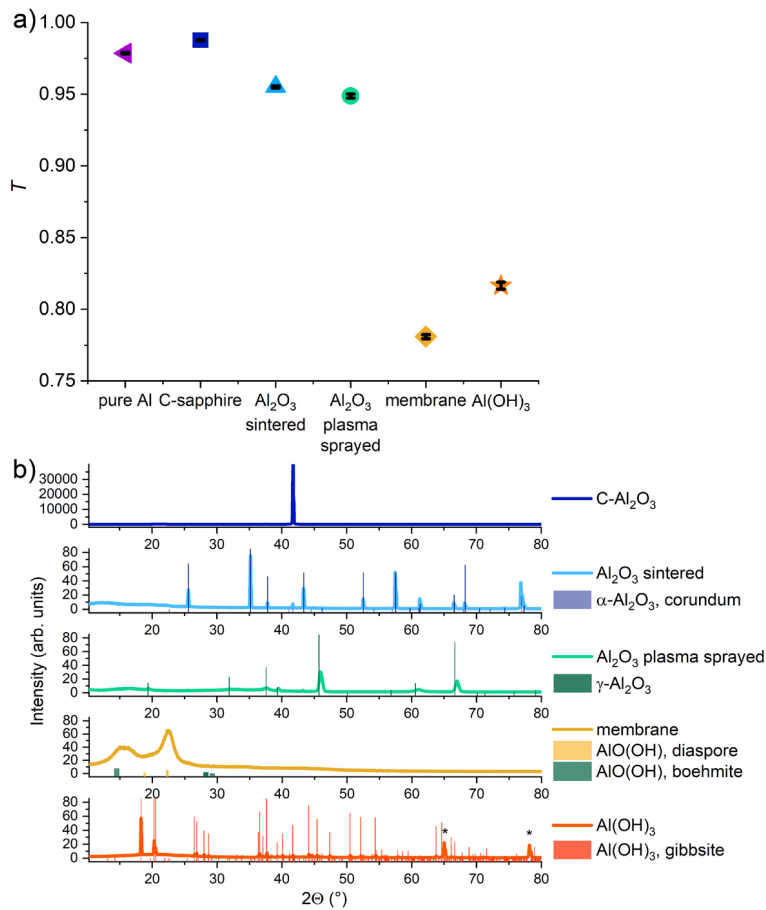

Figure 6: Reference sample characterization: a) averaged transmission value and b) XRD analysis of: C-sapphire, sintered Al oxide, plasma sprayed Al oxide, membrane and Al(OH)$_3$. In Figure a), the transmission of pure Al is added for comparison.

The aluminum oxide grown by plasma spraying consists of metastable, cubic γ-Al$_2$O$_3$. Its transmission value of 0.948 is slightly lower compared to the stable α-Al$_2$O$_3$ phase. This lower transmission value can be attributed to the even more defective nature of this oxide layer, which contains larger structural defects, such as cracks and pores, defects inherent to the plasma spray process and that can integrate hydrogen contamination and/or water in them. These results show, on the other hand, that the changes in transmission value allow to differentiate crystalline Al$_2$O$_3$ based on their crystalline features. Interestingly, the two polycrystalline (with different structures) Al oxides have similar contrast transmission values compared to the porous anodic Al oxides investigated. The Al hydroxide reference sample has a transmission value of 0.816, which is, as expected, lower than the polycrystalline Al oxides. It is worth mentioning that the sample preparation preserved the gibbsite crystalline structure.

Since the porous anodic oxide layers investigated are amorphous, a commercial membrane, Anodisc 13, was included as a comparison. The membrane presents a high degree of pore order, suggesting a two-step anodizing formation process (Figure S7). Traces of phosphor detected by energy dispersive X-Ray analysis (EDX) suggests the use of phosphoric acid as electrolyte. The XRD analysis shows that the Al oxide constituting the membrane is mainly amorphous but presents a different individual pattern than the investigated porous anodic Al oxides. Two broad intensity reflections are visible around 2θ values of 15 ° and 22 °. They can be attributed to boehmite and diaspore-like structures while the peaks assimilated to corundum are no longer present, indicating that the membrane solely consists of Al oxihydroxides and its structure is subsequently more hydrated than the porous anodic Al oxides we grew in phosphoric acids. These findings are in line with the low transmission value of 0.780 for the membrane. It however needs to be mentioned that this value is still unexpectedly lower than gibbsite and does not reflect the structural difference in hydrogen content. This low transmission value may be related to the growth conditions with the long-range pore order facilitating the incorporation of water in the oxide structure during the growth of the membrane.

The investigations on Al oxides/hydroxide reference samples clearly show that the neutron imaging data cannot be solely interpreted based on theoretical cross-section values of pure materials and that both structural and morphological parameters play an important role. This further evidences that neutron imaging is very useful to identify small differences in oxide structure and defects but it also points out that the use of a quasi-reference like the one defined in this study is necessary for data quantification.

## 4. Discussion

These results demonstrate the applicability of high-resolution neutron imaging to study porous anodic Al oxide layers. In particular, due to the high sensitivity of neutrons to hydrogen while $Al_2O_3$ (single crystalline) and electrolyte species are almost neutron transparent, this technique is ideal to visualize and quantify small changes in the water present in porous films as a function of the anodizing conditions.

This preliminary study confirms that porous anodic Al oxides contain water, either as water/hydrogen incorporated in the oxide structure or/and hydration water in the pores. However, the amount of morphological and structural water strongly depends on the anodizing conditions and intrinsically on the oxide formed. Galvanostatic anodizing of pure Al in 0.1 M phosphoric acid leads to the growth of oxides that present similar morphology and amorphous structure up to 6900 s of anodizing, which corresponds to a thickness of 130 µm. The oxide formed after 2320 s (50 µm) however contains two times more water than after 6900 s (130 µm) and present a different water distribution, notably with water enrichment in the top layer constituted of ordered pores. Pore ordering seems therefore to facilitate water incorporation in the pores. Longer anodizing time, for instance 9300 s (230 µm), leads to the formation of an oxide ten times more hydrated than after 6900 s (130 µm). This increase in water content corresponds to a change in the oxide structure associated to the presence of boehmite-like phase instead of gibbsite-like structure. This lower transmission value does not reflect the structural difference in hydrogen content between boehmite and gibbsite but might be attributed to a higher degree of structural disorder as well as a different density of corundum-like and diaspore-like structures compared to thinner oxides. The presence of boehmite could also facilitate incorporation of water in the oxide structure during growth. On the other hand, despite different "amorphous" oxide structure and higher structural disorder, the porous films formed on AW1050 shows similar water content near the air/oxide interface than the ones formed on pure Al whereas notable changes can be observed near the oxide/Al substrate. It seems that anodizing on AW1050 leads to more water entrapment during pore growth at the bottom part of the layers, probably due to higher pore disorder and blockage. Furthermore, at equivalent thickness, the porous anodic Al oxides grown in 0.5 M sulfuric acid contains at least two and half times more water than films grown in 0.1 M phosphoric acid. Although both oxides have similar "amorphous" XRD fingerprint, different morphologies are obtained with regards to the electrolyte used. Anodizing in sulfuric acid leads to the formation of long-range ordered pores with an average size of 20 nm whereas in phosphoric acid, branched pores with a pore diameter of 220 nm are formed. This long-range arrangement of 20 nm pores seems to favor water entrapment in the pores but the difference in the contrast transmission value might also be partly associated to the more defective nature of the oxides formed in sulfuric acid compared to oxides formed in phosphoric acid, as it is the case for crystalline Al oxides. These preliminary results show nevertheless that the different water contents are not necessarily correlated to a specific structural or morphological change but strongly depend on anodizing conditions and therefore, the oxide formed. In a next step, it would be interesting to explore to which extent the changes in structural and morphological water content relate to the intrinsic properties (mechanical, optical and electrical properties and chemical stability) of porous anodic Al oxides.

This study moreover shows the potential of high-resolution neutron imaging to study Al oxide structures from crystalline to amorphous. The technique proves to be sensitive to structural disorder and crystallographic orientations, allowing to differentiate crystalline $Al_2O_3$ based on their crystalline features. Notably, it is worth emphasizing that each of the Al oxides investigated, crystalline or amorphous, presents different characteristics and are not necessarily interchangeable. Crystalline $Al_2O_3$ therefore cannot be used to describe porous anodic Al oxides. Finding a suitable reference sample for porous anodic Al oxides is therefore challenging and requires compromises. The approach taken in this study was the most pragmatic one. By heating the porous anodic Al oxides without inducing any structural changes in the oxide (detectable by XRD), we prove that water can desorb in the top part of the oxide layers, resulting in a "dryer" porous $Al_2O_3$ that can be used as quasi-reference to extract quantitative information. Immersing the sample in water results in reversible filling of the pores. Interestingly, the bottom part of the oxide layers remains unaffected, which suggests that water is trapped near the oxide/Al substrate interface due to pore blockage. Valuable information about the mobility of water within the porous anodic Al oxides can therefore be gained and used to optimize functionalization of the oxides.

The presence of edge enhancement effects at the interfaces however limit data quantification in a region of around 25 µm on both sides of the interface, which not only results in the loss of valuable information about the oxide water content at the interfaces but also indirectly requires a minimum sample thickness of 50 µm. The edge enhancement effects only became detectable in the last couple of years with the increase of spatial resolution in neutron imaging [76,77]. New methodological approaches are therefore being developed to minimize and even suppress them.

The use of high-resolution neutron imaging for the characterization of porous anodic Al oxides offers new perspectives to investigate the structural and morphological water content and distribution in relation to growth conditions, functionalization and stability of the oxides. The fact that for crystalline Al oxides the contrast transmission value can be correlated to the crystallite density and orientation further demonstrates promising application of high-resolution neutron imaging to study the effect of the alloying elements and intermetallic particles on fine variation of the oxide structure formed. With the fast development and advances of high-resolution neutron imaging, higher lateral resolution will soon be achieved. Neutron imaging can then become a very adequate technique to monitor the structural changes occurring in the oxides during application, notably with regards to the formation of hydroxides that can be highly detrimental to surface protection.

## 5. Conclusions

For the first time, to our best knowledge, high-resolution neutron imaging was used to study porous anodic Al oxides. Due to the high sensitivity of neutrons to hydrogen, this technique is very appropriate for the assessment of the water content and distribution in porous anodic Al oxides as a function of anodizing conditions. While single crystal oriented $Al_2O_3$ is almost neutron transparent, we confirm that, as would be expected, porous anodic Al oxides contains structural and morphological water. Variations of overall water content in porous films are not necessarily correlated to a specific structural or morphological change but strongly depend on the anodizing parameters and intrinsically on the oxide formed. High-resolution neutron imaging proves to be highly sensitive to small changes in the water content, which would be highly challenging to characterize with other techniques. We evidence that morphological water desorption in the top part of the oxides can be promoted by heating the samples while subsequent immersion in water, leads to reversible filling of the pores. No changes is observed near the oxide/Al substrate interface, suggesting water entrapment due to pore blockage. This consequently motivates a proper quantification of the water content when discussing the nature and structure of anodic Al oxides in relation to their intrinsic properties and stability. The use of high-resolution neutron imaging is however not limited to porous anodic Al oxides. It shows promising application to characterize crystalline Al oxides, since the contrast transmission value can be associated to the presence of defects, both structural and morphological, in the oxides. Lateral resolution permitting, high-resolution neutron imaging can become the ideal technique to study the changes occurring in the oxides, in particular with regards to the formation and thermal stability of hydroxides, which can be highly detrimental to surface protection and template applications.

**Acknowledgements**

The authors kindly thank Alexander Schmid for his help, since the anodizing parameters (electrolyte concentration, anodizing time) used in this paper were optimized during his project. We are also grateful to Laura Conti for the BET and MIP measurements, Miriam González-Castaño for her contribution during the neutron beamtime 20170107 and Laurence Brassart for the fruitful discussions. This work is based on experiments performed at the Swiss spallation neutron source SINQ, Paul Scherrer Institute, Switzerland.

This research was financed internally by Empa and did not receive any specific grant from funding agencies in the public, commercial, or not-for-profit sectors.

**Author contributions**

**Noémie Ott:** Methodology, Investigation, Validation, Data curation, Writing - Original Draft, Visualization. **Claudia Cancellieri:** Validation, Investigation, Data curation, Writing - Review & Editing, Project administration. **Pavel Trtik:** Methodology; Validation; Resources; Writing - Review & Editing. **Patrik Schmutz:** Conceptualization; Validation; Resources; Writing - Review & Editing; Project administration

**Data availability**

The raw data required to reproduce the findings associated to Al oxides and hydroxide reference samples (XRD and high-resolution neutron imaging) are available to download from [10.5281/zenodo.3944610](10.5281/zenodo.3944610). The raw/processed data required to reproduce the findings associated to porous anodic Al oxide layers cannot be shared at this time as the data also forms part of an ongoing study.

*Supplementary Information*

# High-resolution neutron imaging: a new approach to characterize water in anodic aluminum oxides

Noémie Ott[a,*], Claudia Cancellieri[a], Pavel Trtik[b], Patrik Schmutz[a]

[a] Laboratory for Joining Technologies and Corrosion, Empa, CH-8600 Dübendorf, Switzerland

[b] Neutron Imaging and Activation Group, Laboratory for Neutron Scattering and Imaging, Paul Scherrer Institut, CH-5232 Villigen PSI, Switzerland

* Email: noemie.ott@empa.ch

1. **Neutron imaging – setup**

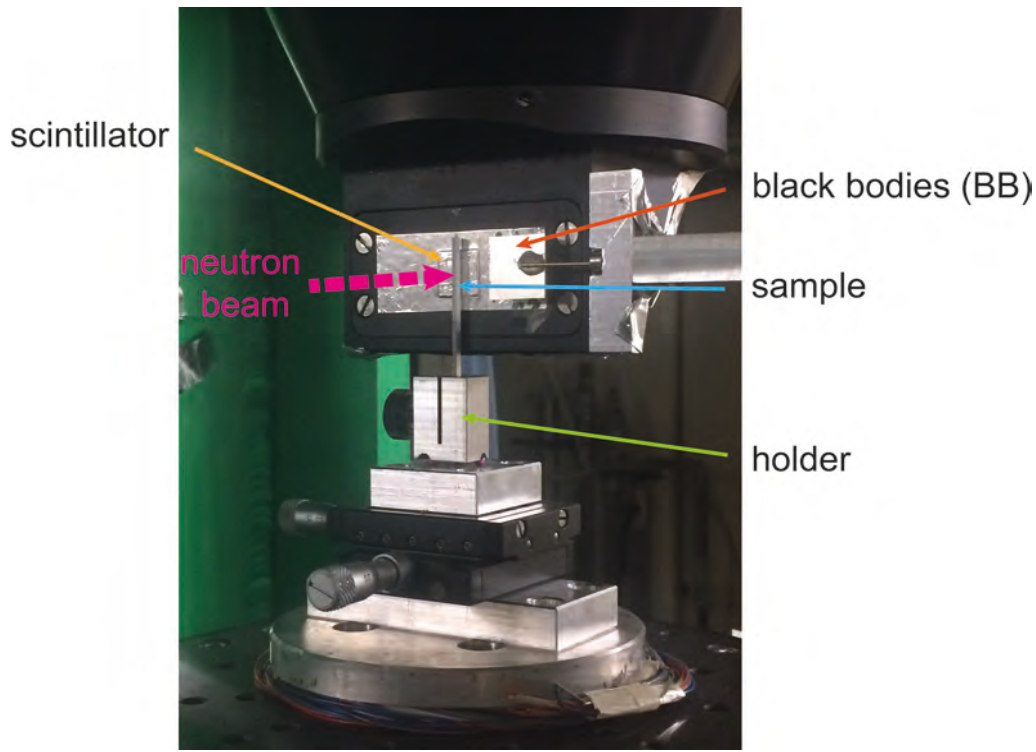

**Figure S1:** Neutron Microscope setup used for neutron imaging. Considering the dimensions of our samples, a custom-made holder had to be designed for these experiments.

**Glossary:**

open beam: no sample in the beam
dark current: background noise without the neutron beam

2. **Neutron imaging – data processing**

All the image processing and quantitative analyses were performed using ImageJ software. For each sample, a stack of 100 projection images was acquired. In addition, stacks of at least 25 open beam (OB) images and open beam images with an interposed frame containing neutron absorbers, black bodies (BB), as well as a stack of 100 images from the sample with the BB were taken for image normalization. The exposure time was 30 s per radiograph. Each stack of images was averaged to form an individual image, which was then normalized using the black body correction procedure developed by the Neutron Imaging and Activation Group (NIAG) at PSI [1,2], described hereafter.

1. Forming the averaged image.

Each stack of images is averaged over the entire stack dimension. Four images per sample are subsequently obtained: averaged open beam (OB) image $I_{OB}$, averaged open beam image with black bodies frame (OB+BB) $I_{OB+BB}$, averaged sample image $I_{sample}$ and averaged sample image with BB frame (sample+BB) $I_{sample+BB}$.

2. Dark current correction

A stack of dark current images was acquired at the beginning of the beamtime. The intensity value obtained was averaged and the mean value determined. This value was then subtracted from the intensity value of $I_{OB}$, $I_{OB+BB}$, $I_{sample}$ and $I_{sample+BB}$ to correct for the background noise.

3. Removing the bright outliers

Outliers – bright pixels due to scattered γ-rays – were removed.

4. Image normalization using BB correction

This has been performed using the "Extended Image Referencing" plugin in ImageJ, developed by the Neutron Imaging and Activation Group (NIAG) at PSI [1-3]. The resulting image $T$ was used for quantitative analyses.

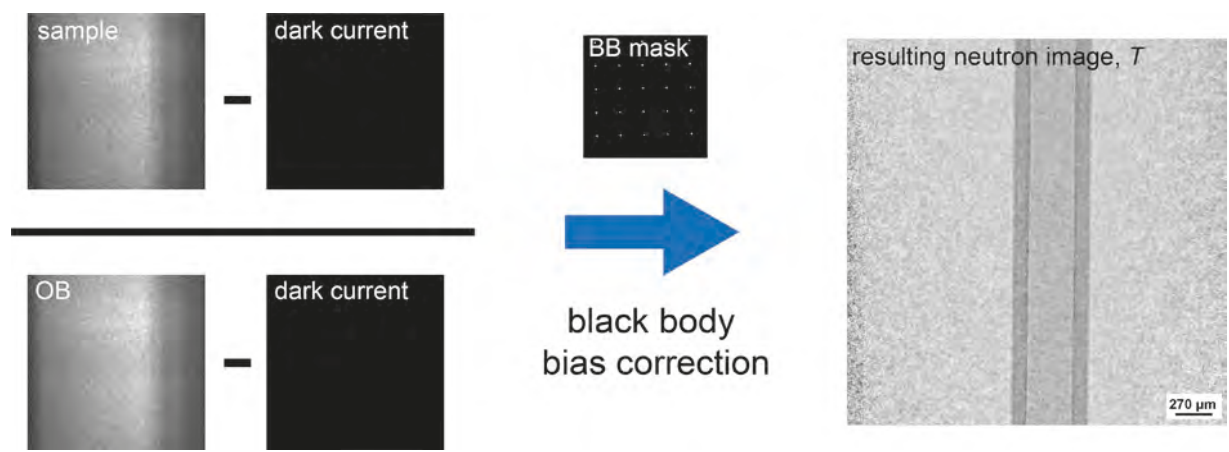

**Figure S2:** Summary of the data processing with the example of pure Al P130

## 3. Determining the linear attenuation coefficient

The attenuation of radiation passing through matter follows the Beer-Lambert law. The transmission contrast intensity $T$ can be expressed as:

$$T = e^{-\Sigma_{tot} d} \quad (1)$$

where $\Sigma_{tot}$ is the linear attenuation coefficient of the material [cm$^{-1}$] through $d$, the sample thickness in the beam direction equal to 2.05 ± 0.03 mm for all studied samples.

The linear attenuation coefficient for a material can be derived experimentally from (1) or it can be calculated using (2) and (3):

$$\Sigma_{tot} = N (\sigma_a + \sigma_s) \quad (2)$$

$$N = N_A \frac{\rho}{M} \quad (3)$$

in which, $\sigma_a$ and $\sigma_s$ are the absorption and scattering microscopic cross sections [cm$^2$], $N$ is the number density [cm$^{-3}$], $N_A$ is the Avogadro number 6.022 10$^{-23}$ [mol$^{-1}$], $\rho$ is the material density [g cm$^{-3}$] and $M$ is the molecular weight [g mol$^{-1}$].

Table S1 compares the values derived from Eq. 1, 2 and 3 with the values measured by neutron imaging, considering: $\rho_{Al}$ = 2.7 g cm$^{-3}$, $\rho_{H_2O}$ = 1.0 g cm$^{-3}$, $\rho_{Al_2O_3,\alpha}$ = 4.0 g cm$^{-3}$, $\rho_{Al(OH)_3}$ = 2.4 g cm$^{-3}$, $M_{Al}$ = 26.98 g mol$^{-1}$, $M_{H_2O}$ = 18.01 g mol$^{-1}$, $M_{Al_2O_3}$ = 101.96 g mol$^{-1}$ and $M_{Al(OH)_3}$ = 78.00 g mol$^{-1}$.

The microscopic cross sections are taken from the NIST Center for Neutron Research[4]: $\sigma_{Al,a}$ = 2.31 10$^{-25}$ cm$^2$, $\sigma_{Al,s}$ =1.503 10$^{-24}$ cm$^2$, $\sigma_{O,a}$ = 1.9 10$^{-28}$ cm$^2$, $\sigma_{O,s}$ =4.232 10$^{-24}$ cm$^2$, $\sigma_{H,a}$ = 3.326 10$^{-25}$ cm$^2$ and $\sigma_{H,s}$ =8.202 10$^{-23}$ cm$^2$.

**Table S1:** comparison of the contrast transmission $T$ calculated and measured.

|  | $\Sigma_{tot}$ calculated [cm$^{-1}$] | $T$ calculated | $T$ measured |
|---|---|---|---|
| Pure Al | 0.104 | 0.979 | 0.978 |
| H$_2$O, free water | 5.65 | 0.32 | 0.47 |
| Al$_2$O$_3$, corundum | 0.377 | 0.92 | 0.955 |
| Al(OH)$_3$, gibbsite | 4.89 | 0.358 | 0.816 |

Table S1 clearly shows that there is a good agreement between the contrast transmission calculated and measured for pure Al. However, as soon as multi-element compounds are considered, the two values differ due to non-negligible additional scattering effects.

The NEA Java-Based Nuclear Data Information System (JAVIS) database[5], can also be used to derive the linear attenuation coefficient and subsequently, the expected transmission value. The transmission value calculated for pure Al at the POLDI beamline is 0.97, which is in good agreement with the measured value. The simulation was also performed for a polycrystalline α-$Al_2O_3$, using the crystal structure of the collection code 9770, taken from the online ICSD database. The transmission value obtained is 0.948, which is relatively closed from the measured value for the polycrystalline corundum. The NEA could however not be used to derive the transmission values for γ-$Al_2O_3$ (metastable phase) or $Al(OH)_3$.

This shows that we need to rely on experimental referencing to determine the linear attenuation coefficient of Al oxides.

## 4. Microstructural characterization of the porous Al anodic oxide layers

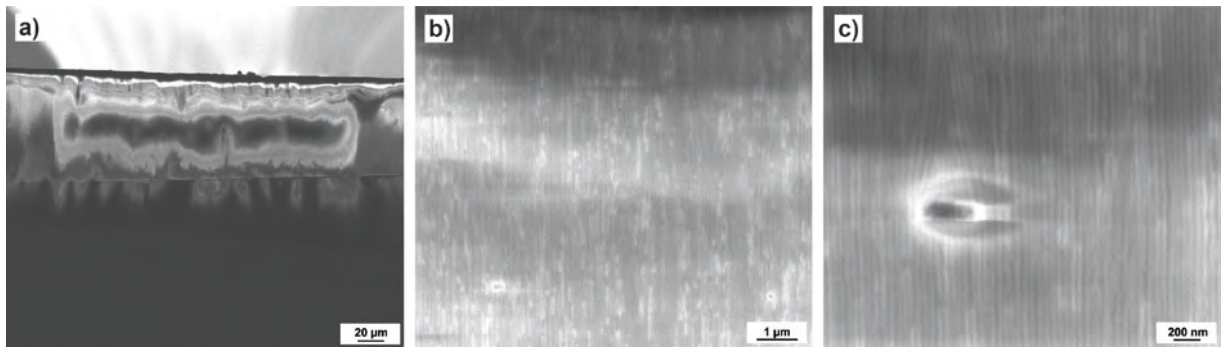

**Figure S3:** SEM images of pure Al S60, porous anodic Al oxide layer formed on pure Al anodized in 0.5 M $H_2SO_4$ at 35 mA $cm^{-2}$ for 3000 s.

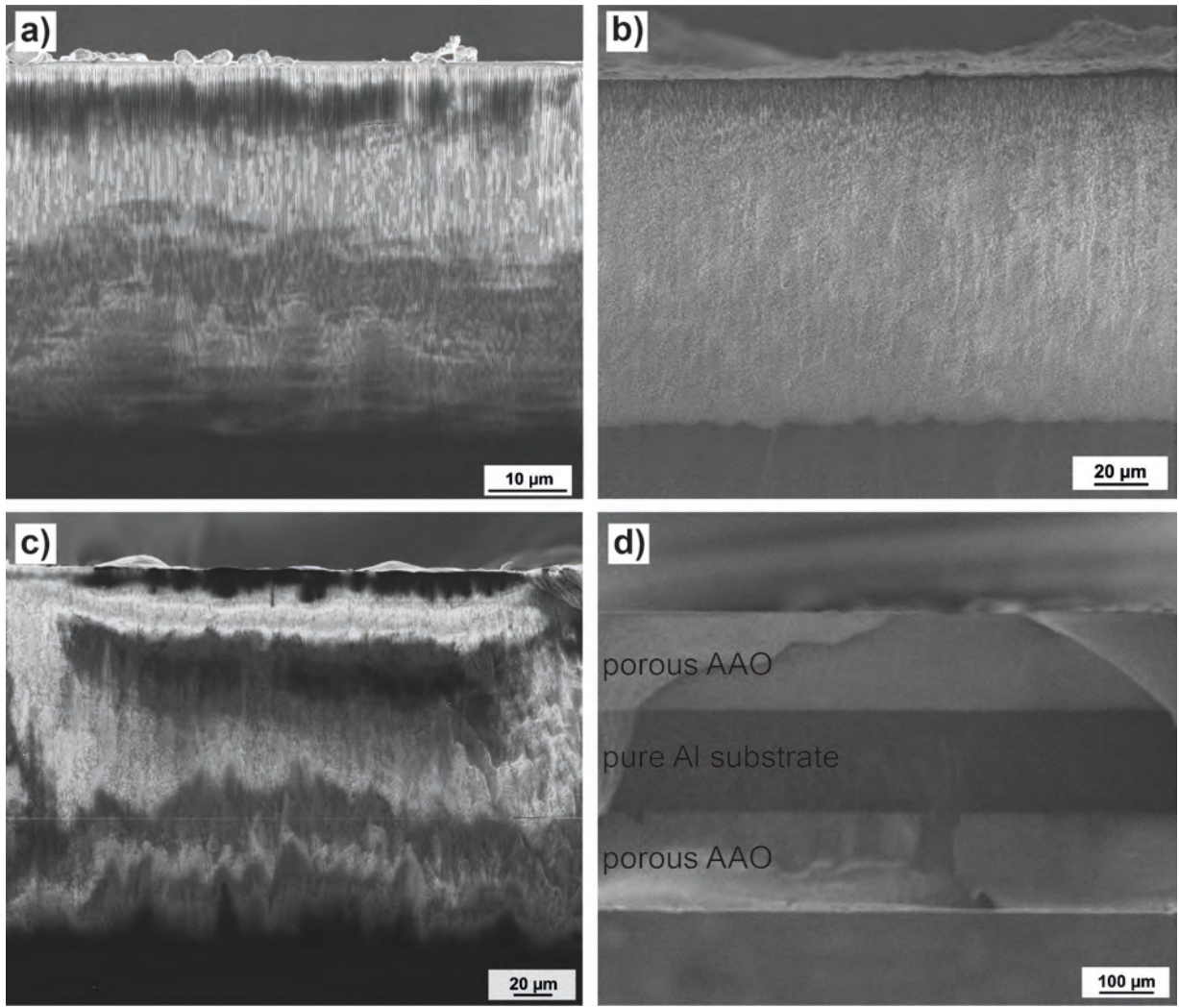

**Figure S4:** SEM images of porous Al anodic oxide (AAO) layers formed on pure Al anodized in 0.1 M $H_3PO_4$ at 35 mA cm$^{-2}$ a) for 2320 s (pure Al P50), b) 6900 s (pure Al P130) and c) 9300 s (pure Al P230); d) low magnification image of pure Al P230, showing both sides of the anodized sample.

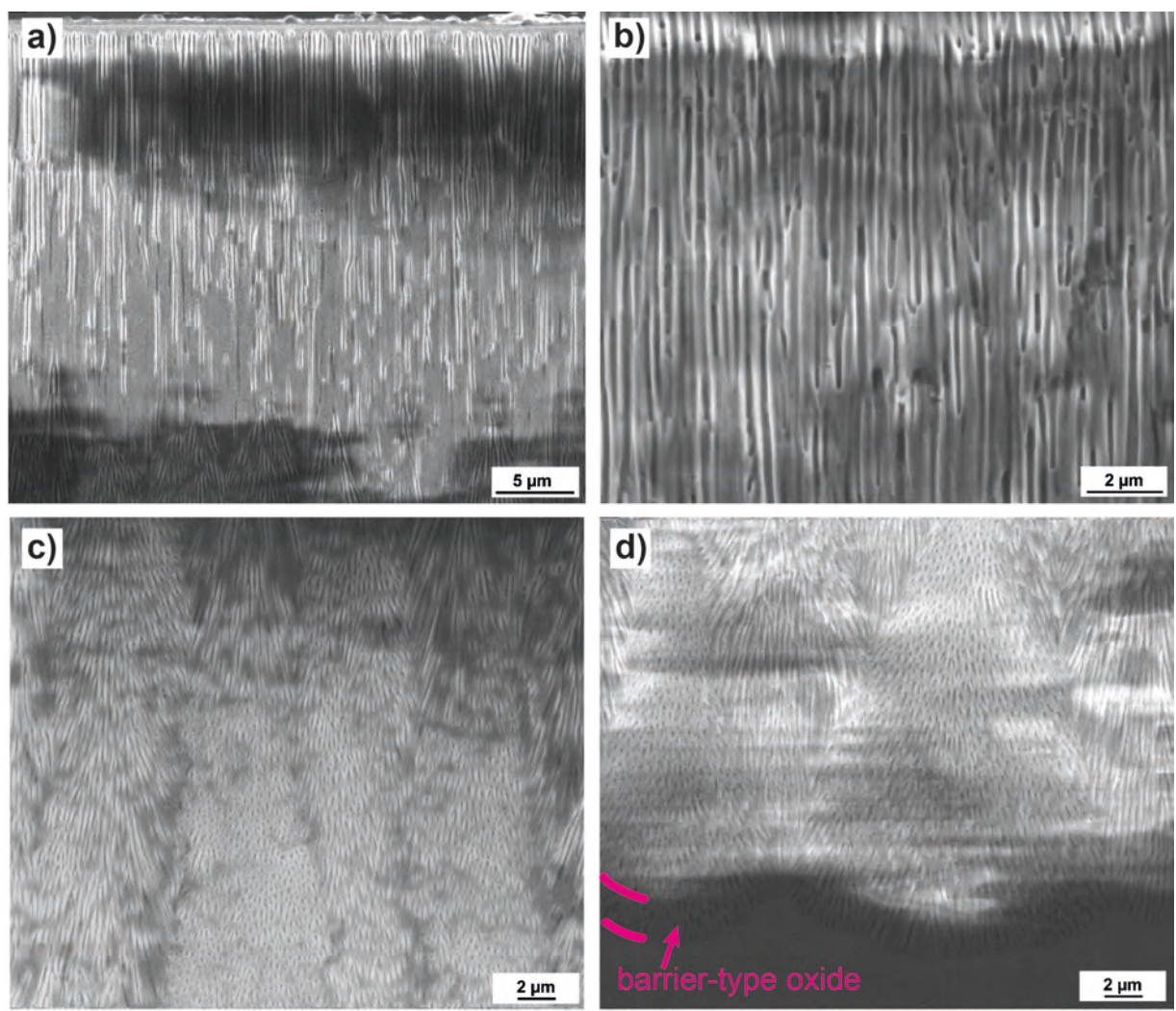

**Figure S5:** Typical SEM images of porous Al anodic oxide (AAO) layers formed on pure Al anodized in 0.1 M $H_3PO_4$ at 35 mA $cm^{-2}$. a) and b) outer part of the Al anodic oxide layer (20-25 μm thick regardless of the anodizing time) showing self-ordering perpendicular to the growth direction; c) branched pores observed in the inner part of the AAO layer; d) porous AAO layer/substrate interface. The barrier-type oxide layer separating the porous AAO layer from the pure Al substrate is enlightened.

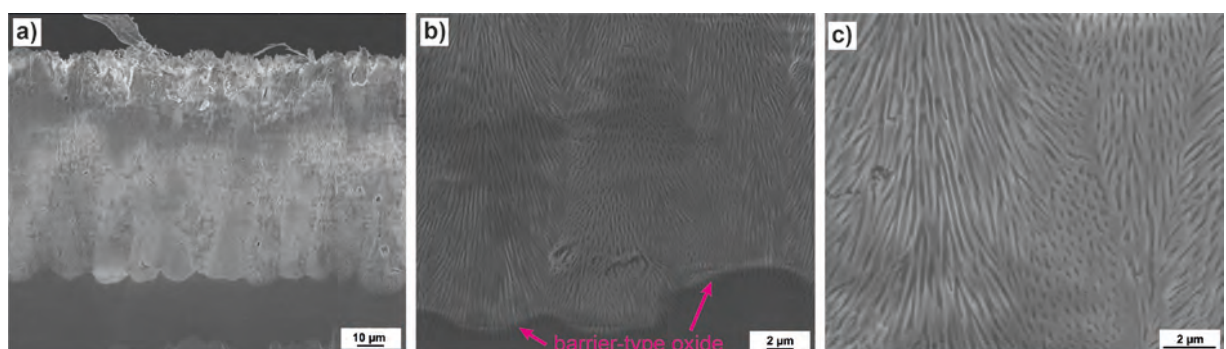

**Figure S6:** SEM images of 1050 P85, porous anodic Al oxide layer formed on AW1050 anodized in 0.1 M $H_3PO_4$ at 35 mA $cm^{-2}$ for 6700 s. In this case, the layer presents low degree of ordering with presence of branched pores across the whole layer thickness.

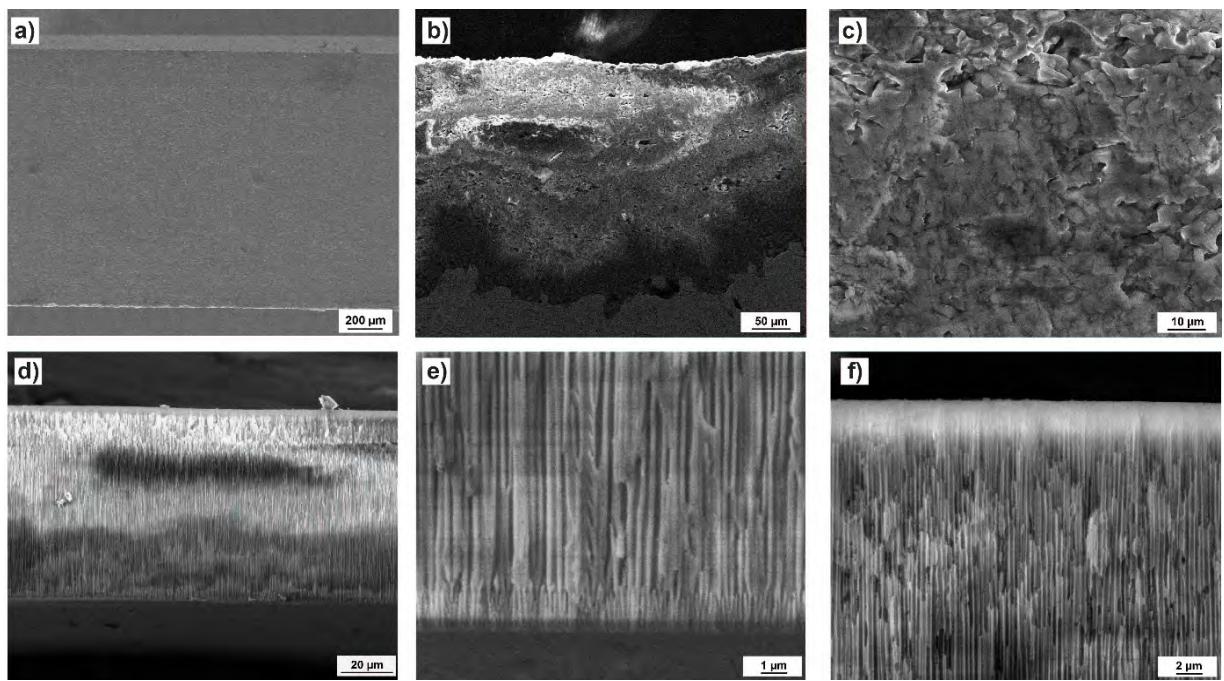

**Figure S7:** SEM images of a) sintered Al oxide, b) and c) plasma sprayed Al oxide, d), e) and f) Al$_2$O$_3$ membrane.

Despite the gold coating, the sample cross sections are charging (Figures S3-S7) but this can hardly be avoided considering the thick Al oxide layers that are investigated. Table S2 reports the thickness and pore sized of the different porous anodic Al oxides investigated. Various reference Al oxides are also included as a comparison. The layer thickness was derived from the neutron images $T$, considering that the size of a pixel corresponds to 2.7 µm, and from the SEM images of the cross-section of the obtained porous Al anodic oxide layers (Figures S3a, S4a, b and c, S6a and S7d). Both values are in good agreement (see Table S2).

The pore size could be roughly estimated from the SEM images. For pure Al S60 (pore size smaller than 100 nm), it was validated by N$_2$ adsorption isotherms at 77 K (Beckmann Coulter SA 3100 surface analyzer), using the Barrett-Joyner-Halenda (BJH) method to derive the average pore size. For the other samples, the pore size was confirmed by mercury intrusion porosimetry (MIP), using a Thermo Pascal 140 and 440 series porosimeter. The pore size of the porous anodic Al oxides formed in 0.1 M H$_3$PO$_4$ are scattered, since the pore growth was not controlled in this study.

**Table S2:** thickness and pore size of the different porous anodic Al oxides investigated.

|  | electrolyte | anodizing time [s] | thickness – SEM [μm] | thickness – neutron images [μm] | pore size [nm] |
|---|---|---|---|---|---|
| pure Al S60 | 0.5 M $H_2SO_4$ | 3000 | 62 ± 6 | 54 ± 3 | 20-25 |
| pure Al P50 | 0.1 M $H_3PO_4$ | 2320 | 50 ± 1 | 45 ± 3 | 220 ± 40 |
| pure Al P130 | | 6900 | 130 ± 1 | 122 ± 6 | 220 ± 40 |
| pure Al P230 | | 9300 | 217 ± 5 | 238 ± 10 | 220 ± 40 |
| 1050 P80 | | 6700 | 80 ± 3 | 83 ± 3 | 230 ± 30 |
| C-$Al_2O_3$ | | | - | 325 ± 10 | - |
| $Al_2O_3$, sintered | | | 995 ± 3 | 1015 ± 40 | - |
| $Al_2O_3$, plasma sprayed | | | 210 ± 20 | 230 ± 25 | - |
| $Al_2O_3$, membrane | | | 78 ± 1 | 77 ± 11 | 315 ± 30 |
| $Al(OH)_3$ | | | - | 54 ± 5 | - |

## 5. XRD characterization

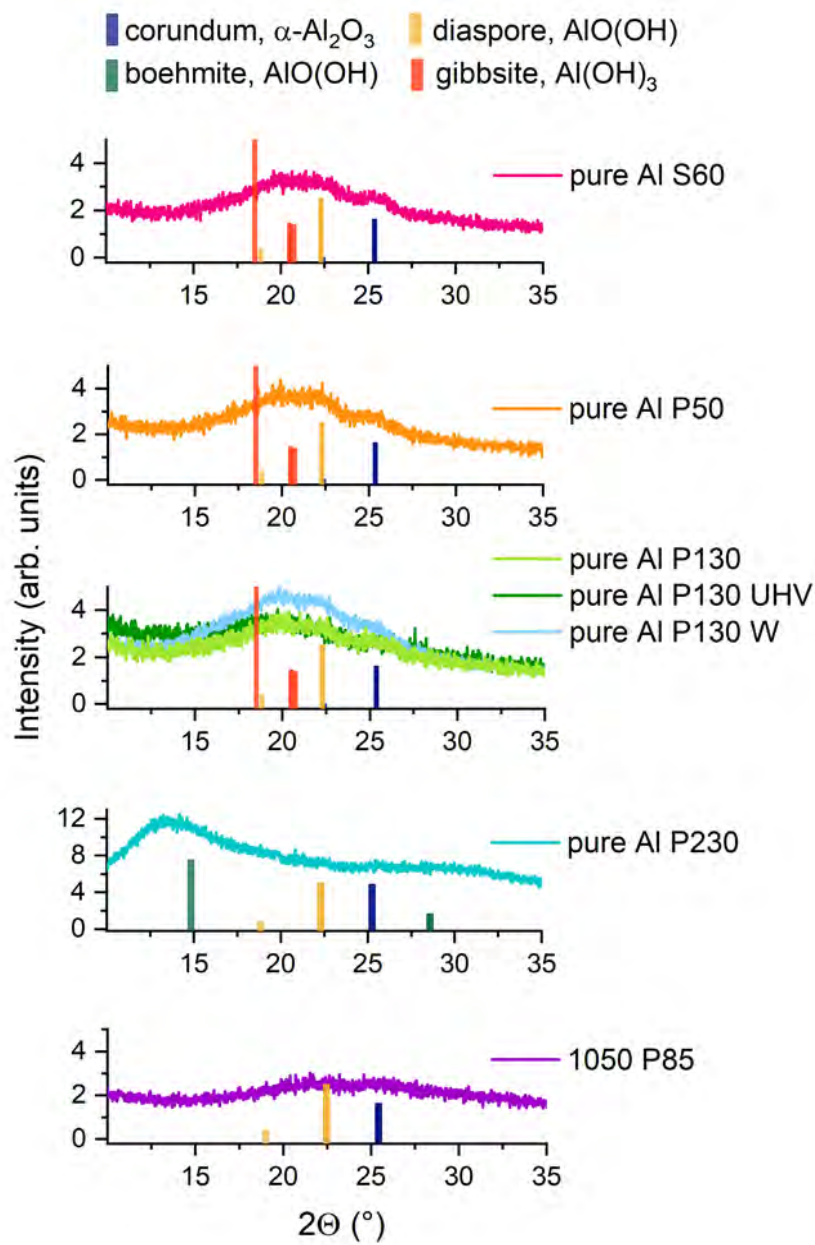

**Figure S8:** Enlargement of XRD scans for pure Al S60, pure Al P50, pure Al P130, pure Al P230 and 1050 P85 in the "amorphous" region and the corresponding peak positions for α-$Al_2O_3$ (corundum – collection code: 9770), AlO(OH) (diaspore – collection code: 29344), $Al(OH)_3$ (gibbsite – collection code: 6162) and AlO(OH) (boehmite – collection code: 27865). These reference powder diffraction patterns were taken from the online ICSD database. The same reference powder diffraction patterns are used in Figure 6.

## 6. Neutron imaging - contrast transmission profiles

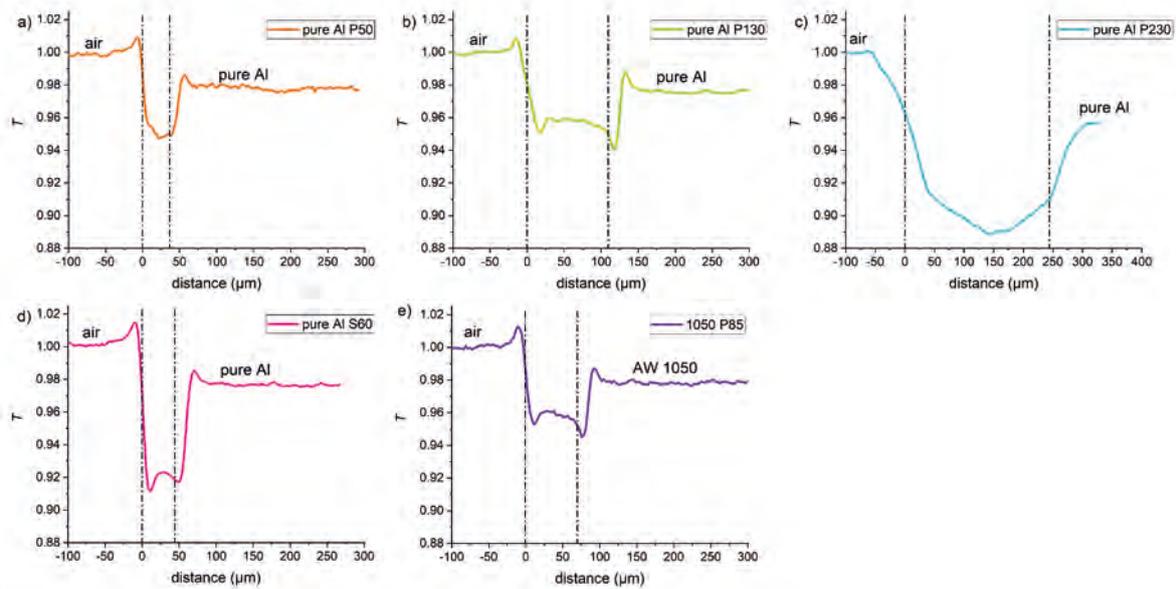

**Figure S9:** Typical contrast transmission profile for a) pure Al P50, b) pure Al P130, c) pure Al P230, d) pure Al S60 and e) 1050 P85. The dashed lines represent the area plotted in Figure 3.

The transmission value $T$ for air is 1.00. Due to instabilities in the neutron beam causing artefacts in the transmission values, this has not been always observed. The profiles presenting discrepancies compared to this value were therefore discarded.